\documentclass[fleqn,10pt]{SelfArx} 

\definecolor{color1}{RGB}{0,0,90} 
\definecolor{color2}{RGB}{0,20,20} 

\usepackage{hyperref} 
\hypersetup{hidelinks,colorlinks,breaklinks=true,urlcolor=color2,citecolor=color1,linkcolor=color1,bookmarksopen=false,pdftitle={Title},pdfauthor={Author},urlcolor=blue}

\usepackage{graphicx}
\usepackage[square]{natbib}
\usepackage{amsmath}
\usepackage{multirow}
\usepackage{subfigure}
\usepackage{pdflscape}

\newcommand{\n}[1]{\mathrm{#1}}

\JournalInfo{Published in Journal of Magnetism and Magnetic Materials, Vol. 322 (21), 3324-3328, 2010} 
\Archive{\href{http://dx.doi.org/10.1016/j.jmmm.2010.06.017}{DOI: 10.1016/j.jmmm.2010.06.017}} 

\PaperTitle{An optimized magnet for magnetic refrigeration} 

\Authors{R. Bj\o{}rk, C. R. H. Bahl, A. Smith, D. V. Christensen and N. Pryds} 
\affiliation{\textit{Department of Energy Conversion and Storage, Technical University of Denmark - DTU, Frederiksborgvej 399, DK-4000 Roskilde, Denmark}} 
\affiliation{*\textbf{Corresponding author}: rabj@dtu.dk} 

\newcommand{\keywordname}{Keywords} 

\Abstract{A magnet designed for use in a magnetic refrigeration device is presented. The magnet is designed by applying two general schemes for improving a magnet design to a concentric Halbach cylinder magnet design and dimensioning and segmenting this design in an optimum way followed by the construction of the actual magnet. The final design generates a peak value of 1.24 T, an average flux density of 0.9 T in a volume of 2 L using only 7.3 L of magnet, and has an average low flux density of 0.08 T also in a 2 L volume. The working point of all the permanent magnet blocks in the design is very close to the maximum energy density. The final design is characterized in terms of a performance parameter, and it is shown that it is one of the best performing magnet designs published for magnetic refrigeration.}

\begin{document}

\flushbottom 

\maketitle 


\thispagestyle{empty} 

\section{Introduction}
Magnetic refrigeration is a potentially energy efficient and environmentally friendly evolving cooling technology that uses the magnetocaloric effect (MCE) to generate cooling through a regenerative process called active magnetic regeneration (AMR).

At present, a great number of magnetic refrigeration test devices have been built and examined in some detail, with focus on the produced temperature span and cooling power of the devices \cite{Barclay_1988, Yu_2003,Gschneidner_2008}. A substantial number of magnet designs have also been published \cite{Coelho_2009,Dupuis_2009,Engelbrecht_2009,Kim_2009,Sari_2009,Tagliafico_2009,Zheng_2009,Tusek_2010} (see Ref. \cite{Bjoerk_2010b} for a review), but for almost all magnet designs no argument is presented for the specific design and dimensioning of the magnet.

In this paper we present the full design approach of a magnet used for magnetic refrigeration. The magnet is designed by applying two general ways or schemes for improving a magnet design to a concentric Halbach cylinder design. The resulting design is dimensioned and segmented and is then characterized by comparing flux density measurements to a numerical simulation. Finally, the magnet design is compared to other magnet designs used in magnetic refrigeration.

\section{Modeling the magnet design}
The magnet is designed for a cylindrical rotating magnetic refrigeration device under construction at Ris\o{} DTU, in which plates of magnetocaloric material rotate in an air gap between an outer and inner cylindrical magnetic structure. The dimensions of the design, which have been chosen based on the desired temperature span and cooling capacity of the device, are such that the volume between the inner and outer magnet is 4 L. The magnetic refrigeration device is designed such that the magnet must provide four high flux density regions and four low flux density regions in the air gap between the two magnets.

A similar magnetic refrigerator, i.e. with a stationary magnet and a rotating magnetocaloric material, was also considered in Ref. \cite{Tusek_2010}, where a magnet design that generates a magnetic field between 0.1 T and 1 T in four low and four high field regions is presented. Rotary magnetic refrigerators where the magnet rotates and the magnetocaloric material is kept stationary are considered in, e.g., Refs. \cite{Vasile_2006}, \cite{Okamura_2007} and \cite{Zimm_2007}. One of these designs uses rectangular magnets while the other two use highly irregularly shaped magnets. The generated magnetic field is between 1.0 and 1.9 T, although the latter value is based on a two dimensional numerical simulation which is known to overestimate the magnitude of the magnetic field except for very long assemblies.

Based on numerical modeling of the AMR process using the model of Ref. \cite{Nielsen_2009a} the length of the Ris\o{} DTU device was chosen to be 250 mm \cite{Nielsen_2010_priv}. Based on practical engineering requirements, as well as to allow ample room for the inner magnet, an external radius of the inner magnet of 70 mm and an internal radius of the outer magnet of 100 mm was chosen.

The regenerator itself can consist of either plates or a packed bed of magnetocaloric material. The dimensions, shape and stacking of the plates or the dimensions and shape of the packed bed can vary, and the performance of the refrigeration device will of course depend on these parameters. The magnetocaloric material is contained in a plastic structure with low heat conduction, so the heat transfer between the magnet and the magnetocaloric material is kept low. As the magnetocaloric material is rotated in the magnetic field there will be an eddy current induced in the magnetocaloric material. The heating due to this eddy current is negligible because the magnetization is small and the rotation rate is only on the order of 1 Hz.

\begin{figure}[!t]
\centering
\includegraphics[width=1\columnwidth]{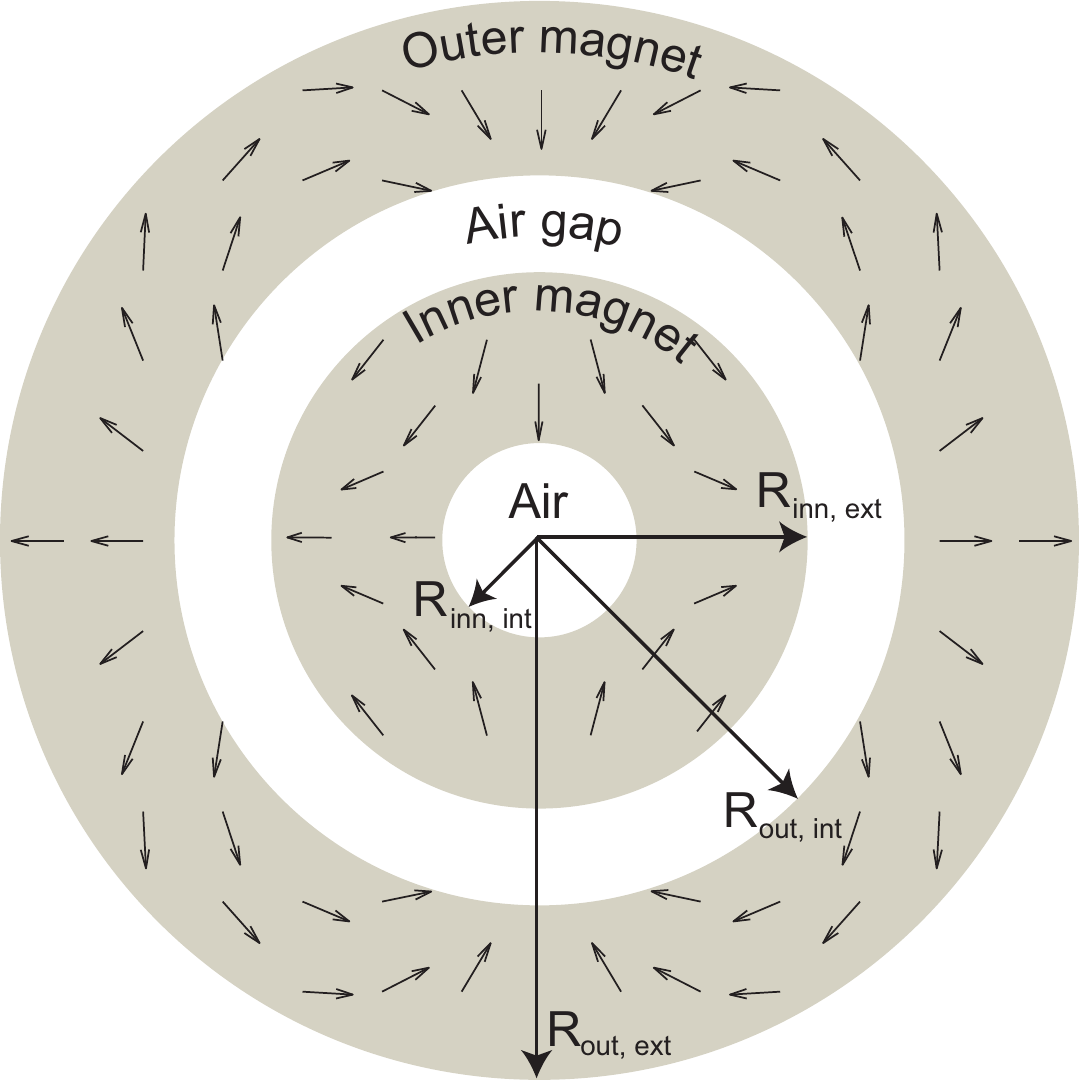}
\caption{The concentric Halbach cylinder design. The direction of magnetization is shown as arrows. The different radii have been indicated.}\label{Fig.Dimensions_drawing}
\end{figure}

A magnet design that fulfils the requirement of generating four high and low flux density regions is the concentric Halbach cylinder design \cite{Bjoerk_2010a}. Here each cylinder is magnetized such that the remanent flux density at any point varies continuously as, in polar coordinates,
\begin{eqnarray}
B_{\mathrm{rem},r}    &=& B_{\mathrm{rem}}\; \textrm{cos}(p\phi) \nonumber\\
B_{\mathrm{rem},\phi} &=& B_{\mathrm{rem}}\; \textrm{sin}(p\phi)\;,\label{Eq.Halbach_magnetization}
\end{eqnarray}
where $B_{\mathrm{rem}}$ is the magnitude of the remanent flux density and $p$ is an integer \cite{Mallinson_1973,Halbach_1980}. Subscript $r$ denotes the radial component of the remanence and subscript $\phi$ the tangential component. A magnet with four high and four low flux density regions, as described above, can be created by having a $p=2$ outer Halbach cylinder and a $p=-2$ inner Halbach cylinder. The concentric Halbach cylinder design is shown in Fig. \ref{Fig.Dimensions_drawing}.

This magnet design is the starting design for the optimized magnet design presented here. The design is improved by applying an algorithm to increase the difference in flux density between a high and low flux density region in an air gap in a magnetic structure, as described in Ref. \cite{Bjoerk_2010c}. The algorithm lowers the flux density in a given area by replacing magnet material enclosed by an equipotential line of the magnetic vector potential, $A_\mathrm{z}$, with a soft magnetic material or vacuum.

Furthermore, the design is improved by replacing magnet material with a high permeability soft magnetic material where the component of the magnetic field along the remanence is not large and negative, i.e. where $\mu_0 \mathbf{H}\cdot{}\mathbf{\hat{B}}_\mathrm{rem} > -\gamma$, with an appropriately chosen positive $\gamma$, as here a high permeability soft magnetic material will produce a similar value of $|\mathbf{B}|$ as the magnet produces \cite{Bloch_1998, Coey_2003, Bjoerk_2010c}.

These two improvements are applied to the design using a numerical two dimensional model implemented in the commercially available finite element multiphysics program \emph{Comsol Multiphysics} \cite{Comsol} and using magnets with a remanence of 1.44 T and a relative permeability of 1.05, which are the properties of standard neodymium-iron-boron (NdFeB) magnets \cite{Standard}. A two dimensional model is used as the magnet design is symmetric along the length of the design and the ratio of the gap to the length of the assembly is much smaller than 1, making end effects relatively unimportant.

For the algorithm the equipotential line of $A_\mathrm{z}$ is chosen to be the line that goes through the point $(r = 100 \;\mathrm{mm}, \phi{} = 22.5^{\circ})$, i.e. the point on the internal radius of the outer magnet, half way between the center of the high and low flux density regions, as this equipotential line encircles the low flux density region. Iron is used as the soft magnetic material as it has a very high permeability as well as being easily workable and reasonable priced. A value of $\gamma=0.125$ T is used for replacing magnet material with iron where the component of the magnetic field along the remanence is not large and negative.

\subsection{Dimensioning of the design}
\begin{figure}[!t]
\centering
\includegraphics[width=1\columnwidth]{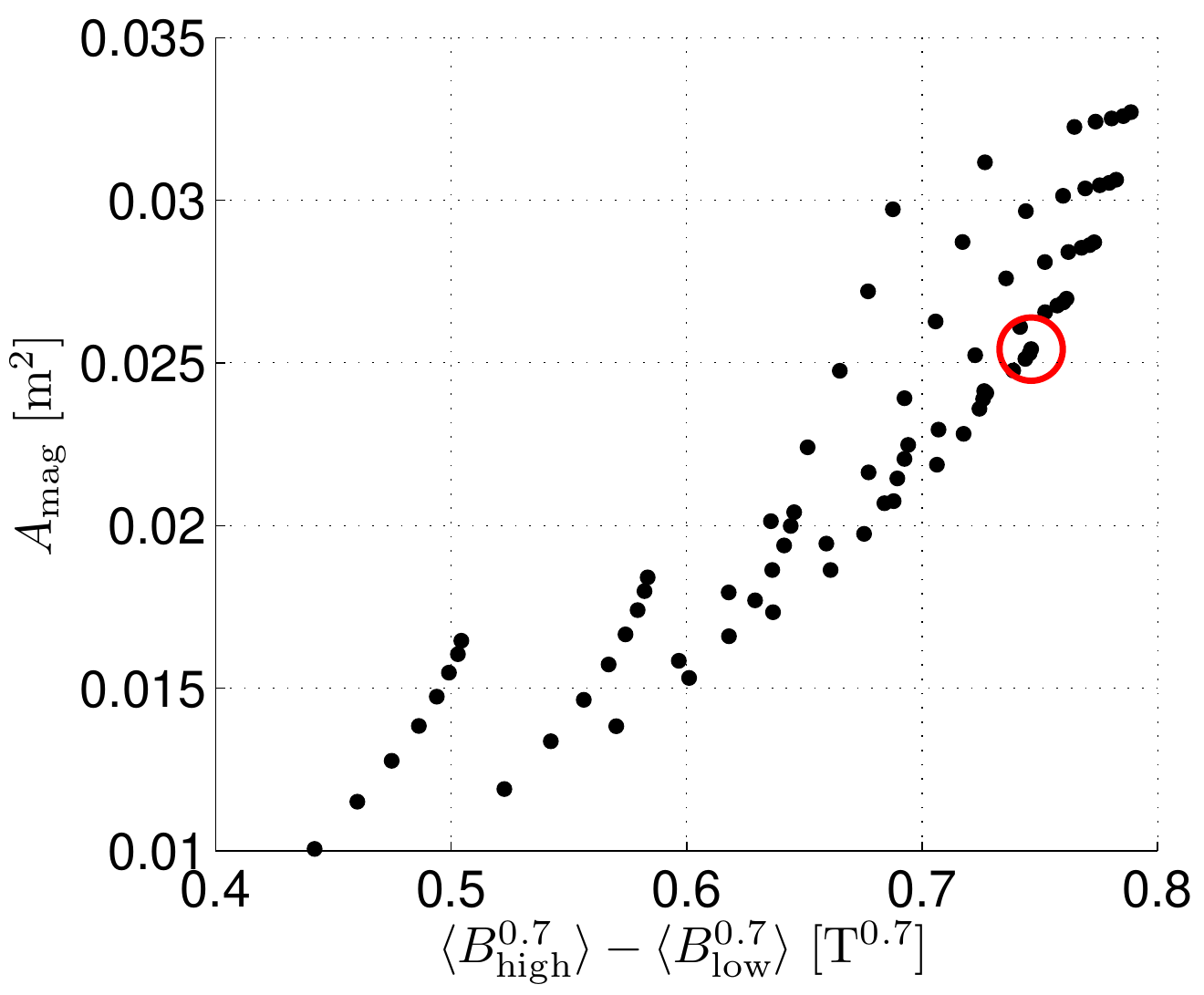}
\caption{(Color online) The difference in flux between the high and low flux density regions to the power of 0.7 as a function of the cross-sectional area of the magnet, $A_\mathrm{mag}$, for a range of different external radii of the outer magnet, $R_\mathrm{out,ext}$, and internal radii of the inner magnet, $R_\mathrm{inn,int}$. The area is used as the model is two-dimensional. The chosen set of dimensions have been encircled.}\label{Fig.Determine_size_of_magnets}
\end{figure}

The remaining dimensions of the magnet design, i.e. the external radius of the outer magnet, $R_\mathrm{out,ext}$, and the internal radius of the inner magnet, $R_\mathrm{inn,int}$, are chosen based on a parameter variation of the concentric Halbach design where the two improvements discussed above have been applied. The external radius of the outer magnet was varied from 110 mm to 155 mm in steps of 5 mm and the internal radius of the inner magnet was varied from 10 mm to 50 mm in seven equidistant steps. The optimization parameter is taken as the difference in flux density between the high and low flux density regions to the power of 0.7 as a function of the cross-sectional area of the magnet; this is shown in Fig. \ref{Fig.Determine_size_of_magnets}. Here $\langle{}B_\n{high}^{0.7}\rangle{}$ denotes the average of the flux density to the power of 0.7 in the high field region and similarly $\langle{}B_\n{low}^{0.7}\rangle{}$ for the low flux density region. For this design the high and low flux density regions are defined to be of the same size and span an angle of 45 degree each making them adjacent.

The reason the power of 0.7 is chosen is that the adiabatic temperature change of a second order magnetocaloric material scales with the magnetic field to the power of $0.7$ at the Curie temperature \cite{Bjoerk_2010b,Pecharsky_2006}. Thus it is this value that is important to the performance of a magnet used in magnetic refrigeration.

To limit the cost of the magnet, a cross-section of $A_\mathrm{mag}$ = 0.025 m$^2$ was chosen. Based on this value and Fig. \ref{Fig.Determine_size_of_magnets} the optimal design was chosen. This design has an external radius of the outer magnet of 135 mm and an internal radius of the inner magnet of 10 mm.

\begin{figure*}[!t]
\centering
\subfigure[Original design]{\includegraphics[width=1\columnwidth]{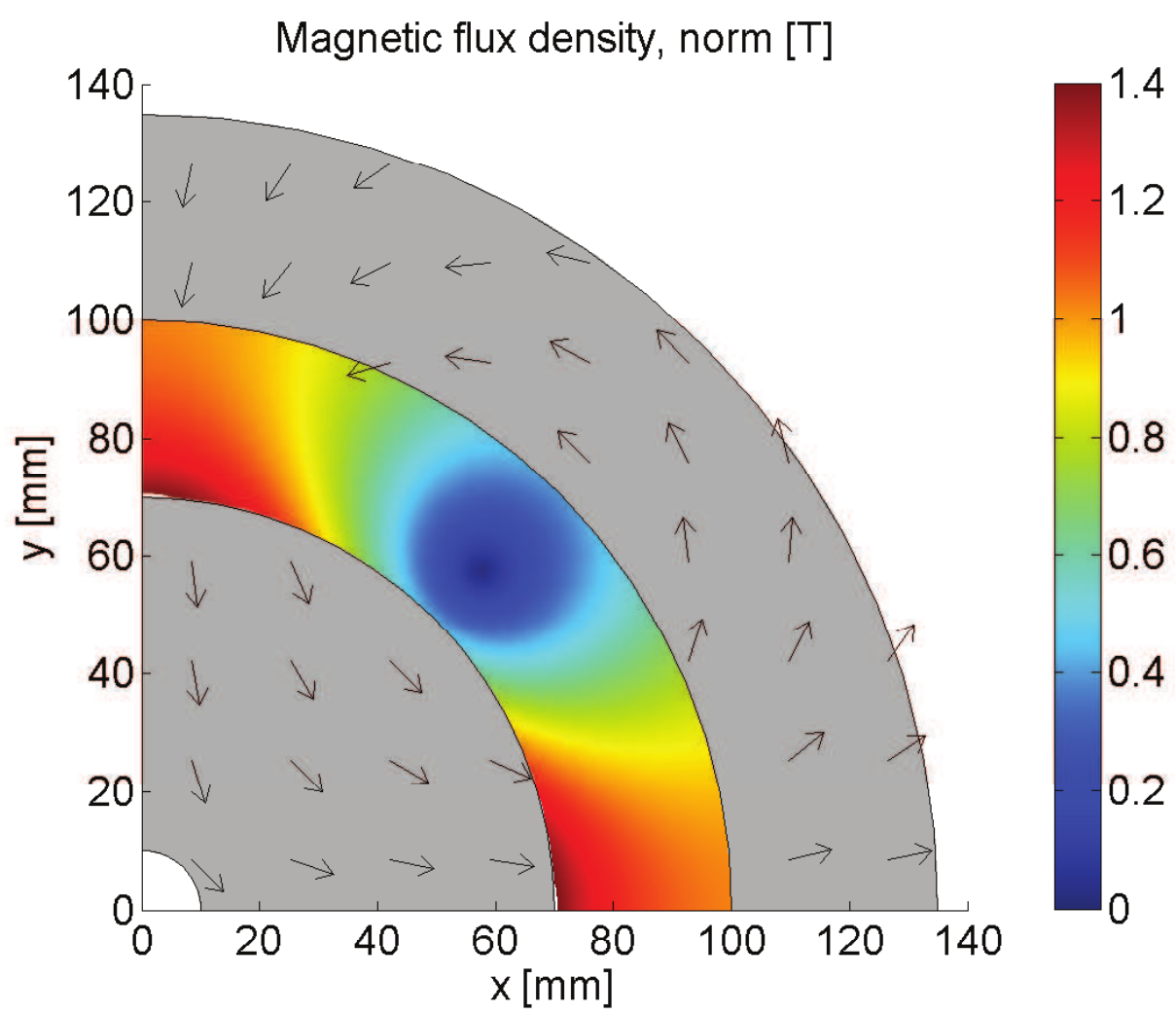}}\hspace{0.4cm}
\subfigure[Design with applied improvements]{\includegraphics[width=1\columnwidth]{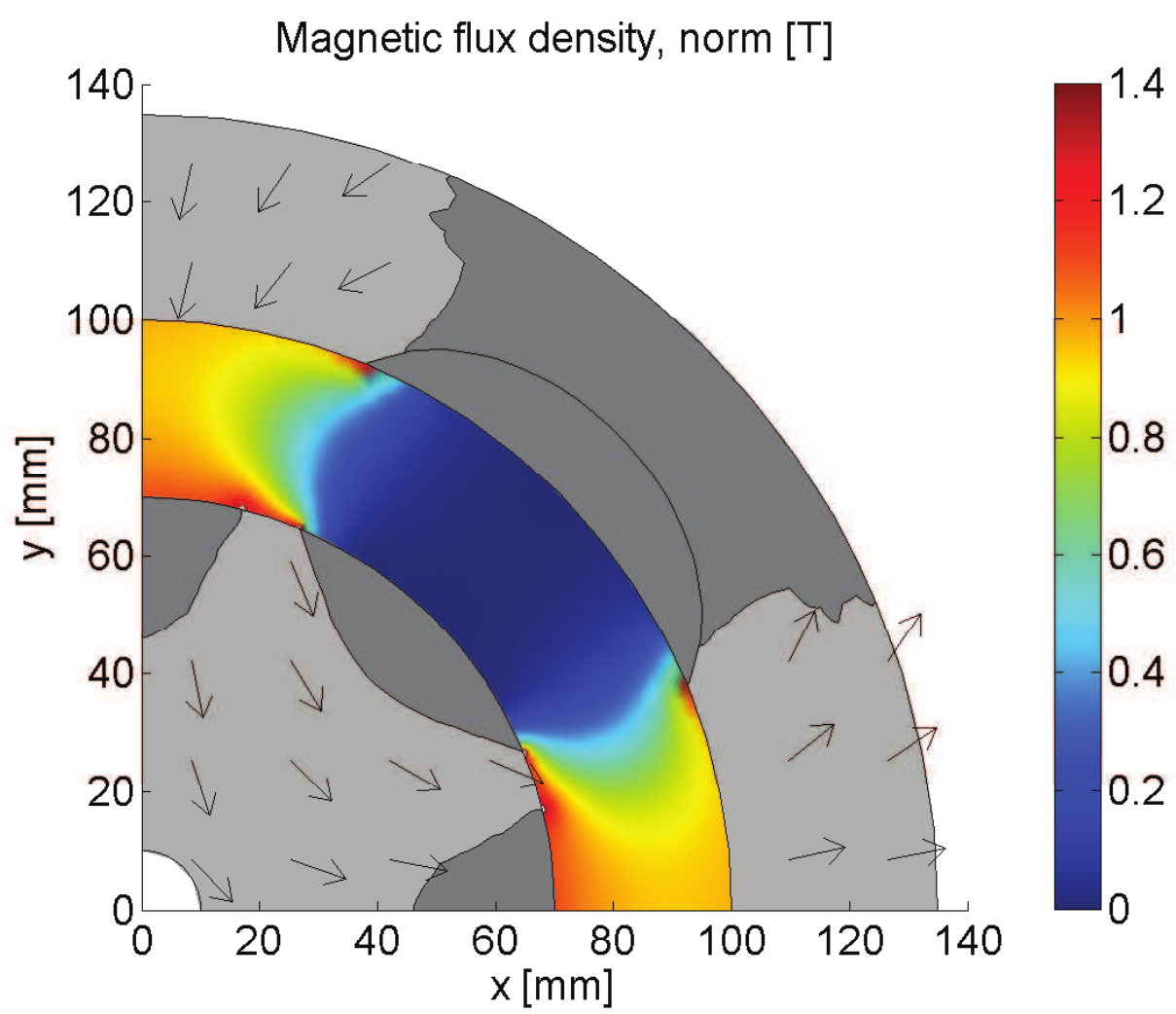}}\\
\caption{(Color online) Fig (a) shows a quadrant of the a concentric Halbach cylinder with $p_\mathrm{outer}=2$ and $p_\mathrm{inner}=-2$. The remaining quadrants can be obtained by mirroring along the coordinate axes. The magnetization is shown as black arrows on the magnets, which are light grey. Iron is dark grey. The flux density in the air gap between the cylinders is shown as a color map. Fig (b) shows the same design after the two improvement schemes have been applied. The line in the iron region in the outer magnet separates the iron regions generated by the two improvement schemes, and it is only shown for reference.}\label{Fig.Opt_magnet}
\end{figure*}

The original concentric Halbach cylinder design and the design after the application of the different improvements are shown in Fig. \ref{Fig.Opt_magnet} for the dimensions found above.

\section{The physical magnet}\label{The physical magnet}

\subsection{Segmentation of the final design}
To allow construction of the magnet, the design shown in Fig. \ref{Fig.Opt_magnet}(b) must be segmented. The number of segments is an important parameter as the more segments used the more expensive the manufacturing process becomes. Generally it is the total number of segments that determines the cost together with the overall magnet volume, due to the handling of the individual segments. However, segments with different geometric shapes introduce an additional cost as these must be separately manufactured. If different segments have the same geometrical shape but different directions of magnetization these introduce little additional cost as the same molds and fixation tools can be used \cite{Flemming_2009}.

\begin{figure}[tb]
\centering
\includegraphics[width=1\columnwidth]{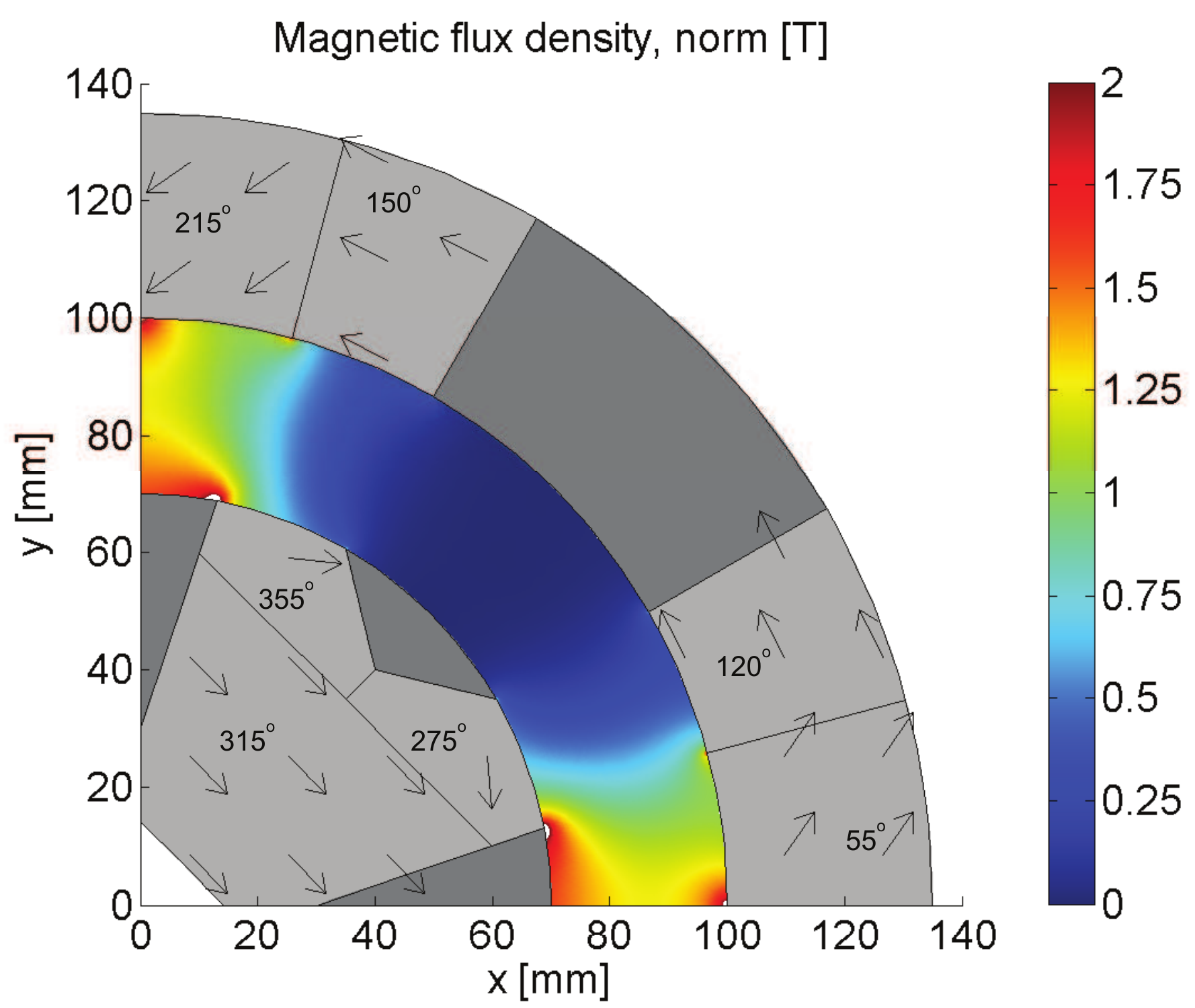}
\caption{(Color online) The segmentation of the final design. The direction of magnetization has been found by maximizing $\langle{}B_\n{high}^{0.7}\rangle{}-\langle{}B_\n{low}^{0.7}\rangle{}$. The direction of magnetization is indicated on each segment. The small white areas in the air gap have a flux density higher than the maximum value on the color bar.}\label{Fig.model_f_1_Optimum_1_2D}
\end{figure}

The segmentation of the optimized design is done manually. The size of the iron regions is decreased a bit in order to generate a higher flux density in the high flux density region. In order to find the optimal direction of magnetization of the individual segments an optimization procedure has been applied. The optimization routine used is a modified version of the \emph{Matlab} function \verb+FMINSEARCH+ \cite{Matlab}, called \verb+FMINSEARCHBND+, which finds the minimum of an unconstrained multivariable function with boundaries using a derivative-free method \cite{fminsearchbnd}. A Comsol model with a predefined geometry is used as input, with the direction of magnetization as variables. The optimization criteria is that the difference between $\langle{}B_\n{high}^{0.7}\rangle{}$ and $\langle{}B_\n{low}^{0.7}\rangle{}$ be maximized. The segmentation of the magnet design and the resulting directions of magnetization are shown in Fig. \ref{Fig.model_f_1_Optimum_1_2D}.

\begin{figure}[tb]
\centering
\includegraphics[width=1\columnwidth]{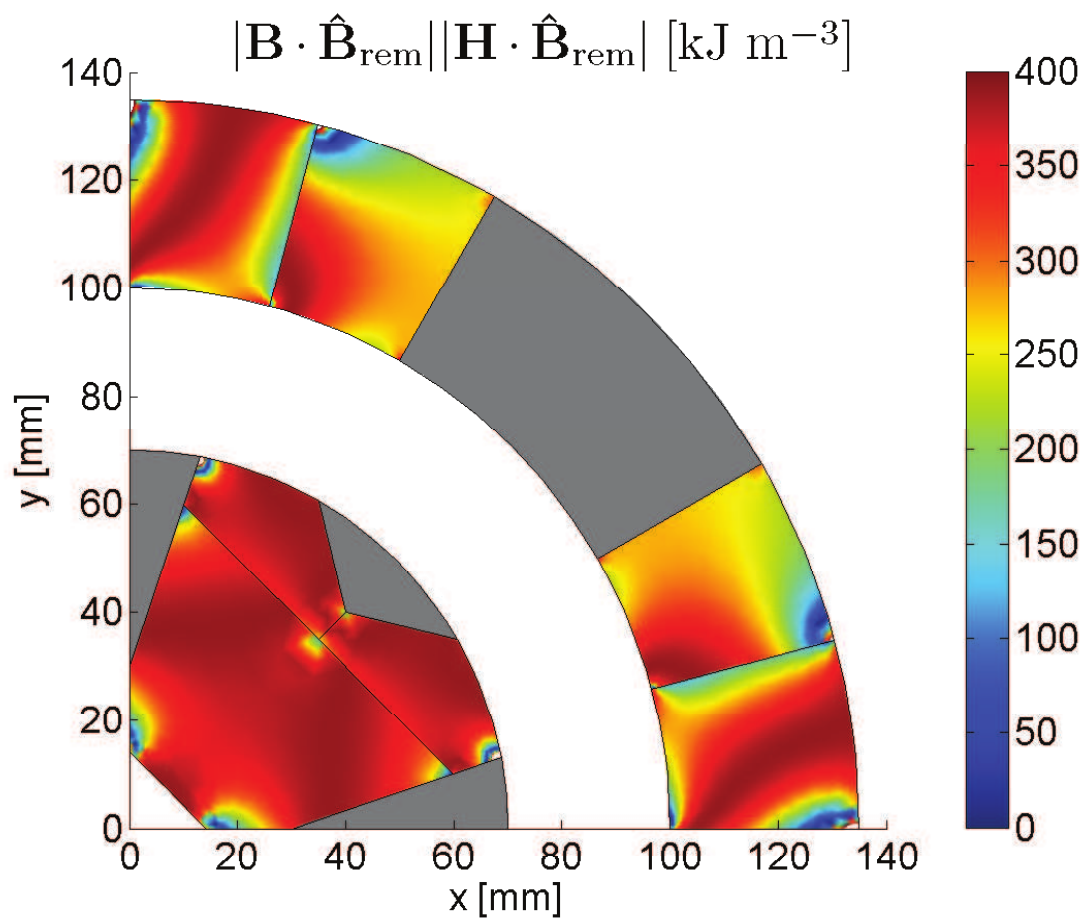}
\caption{(Color online) The working point, $|\mathbf{B}\cdot{}\mathbf{\hat{B}}_\mathrm{rem}||\mathbf{H}\cdot{}\mathbf{\hat{B}}_\mathrm{rem}|$, of the magnets. The maximum working point for a 1.44 T remanence magnet, as is used here, is 400 kJ m$^{-3}$. }\label{Fig.model_f_1_Optimum_1_2D_Working_point}
\end{figure}

The effectiveness of the magnet design can be judged from the working point of the magnets, i.e. the size of the magnetic field times the size of the flux density, both measured in the direction of the remanence: $|\mathbf{B}\cdot{}\mathbf{\hat{B}}_\mathrm{rem}||\mathbf{H}\cdot{}\mathbf{\hat{B}}_\mathrm{rem}|$. In Fig. \ref{Fig.model_f_1_Optimum_1_2D_Working_point} the working point is shown as calculated from a model of the magnet design. For magnets with a remanence of 1.44 T, as is used here, the maximum energy density, i.e. the optimal working point $(|\mathbf{B}\cdot{}\mathbf{\hat{B}}_\mathrm{rem}||\mathbf{H}\cdot{}\mathbf{\hat{B}}_\mathrm{rem}|)_\n{max}$, is 400 kJ m$^{-3}$ \cite{VAC_specifications_2007}. As can be seen from the figure most parts of the magnets are close to the maximum energy density  thus illustrating the efficiency of the design.

\subsection{The final design realized}
\begin{figure}[tb]
\centering
\includegraphics[width=1\columnwidth]{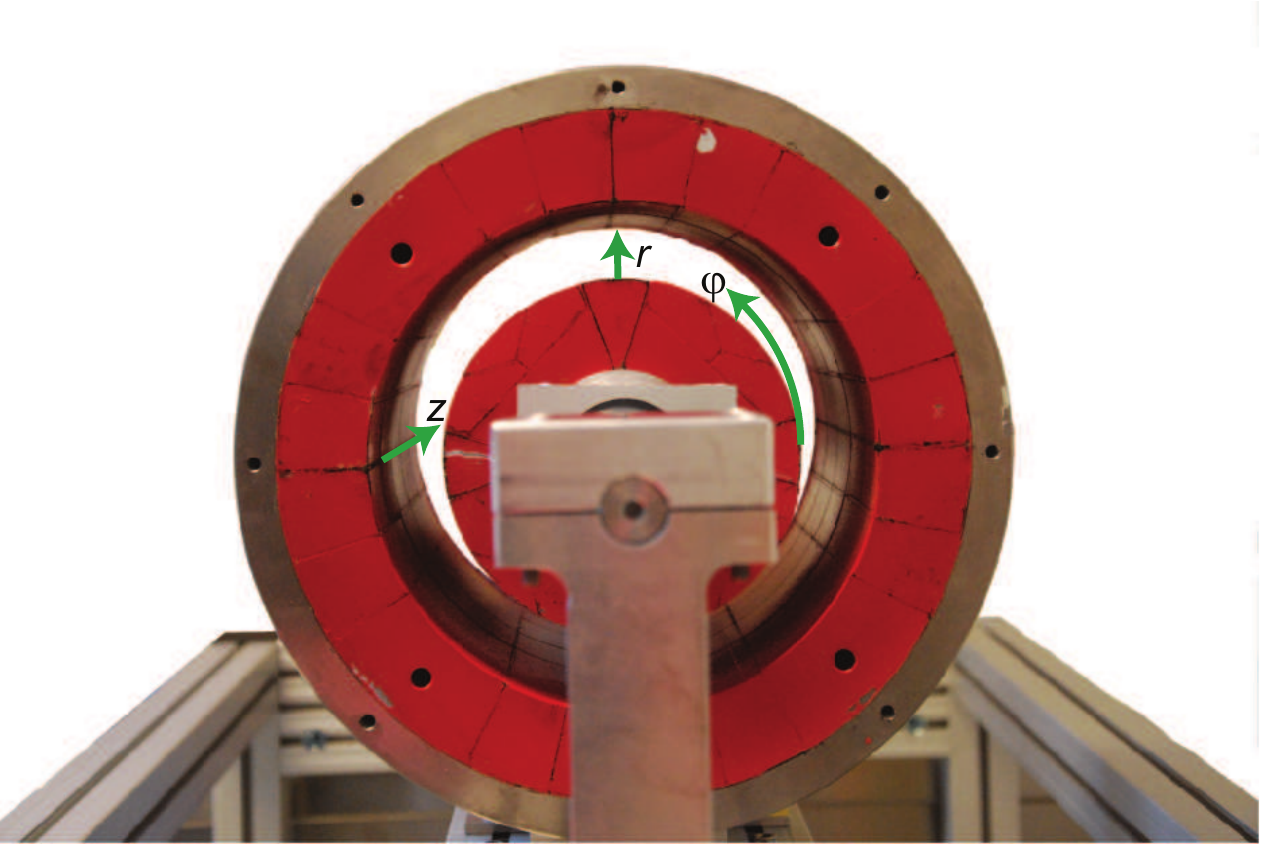}
\caption{(Color online) A photo of the actual constructed magnet (in red) including a stand and an outer stainless steel casing. The coordinate system used for the measurements of the flux density is also shown.}\label{Fig.Magnet_picture}
\end{figure}

The magnet design shown in Fig. \ref{Fig.model_f_1_Optimum_1_2D} has been constructed and a photo of the magnet is shown in Fig. \ref{Fig.Magnet_picture}. The magnet has a length of 250 mm.

\begin{figure}[!t]
\centering
\subfigure[$|B|$ as function of $\phi$ in the middle of the air gap, $r=85$ mm.]{\includegraphics[width=.91\columnwidth]{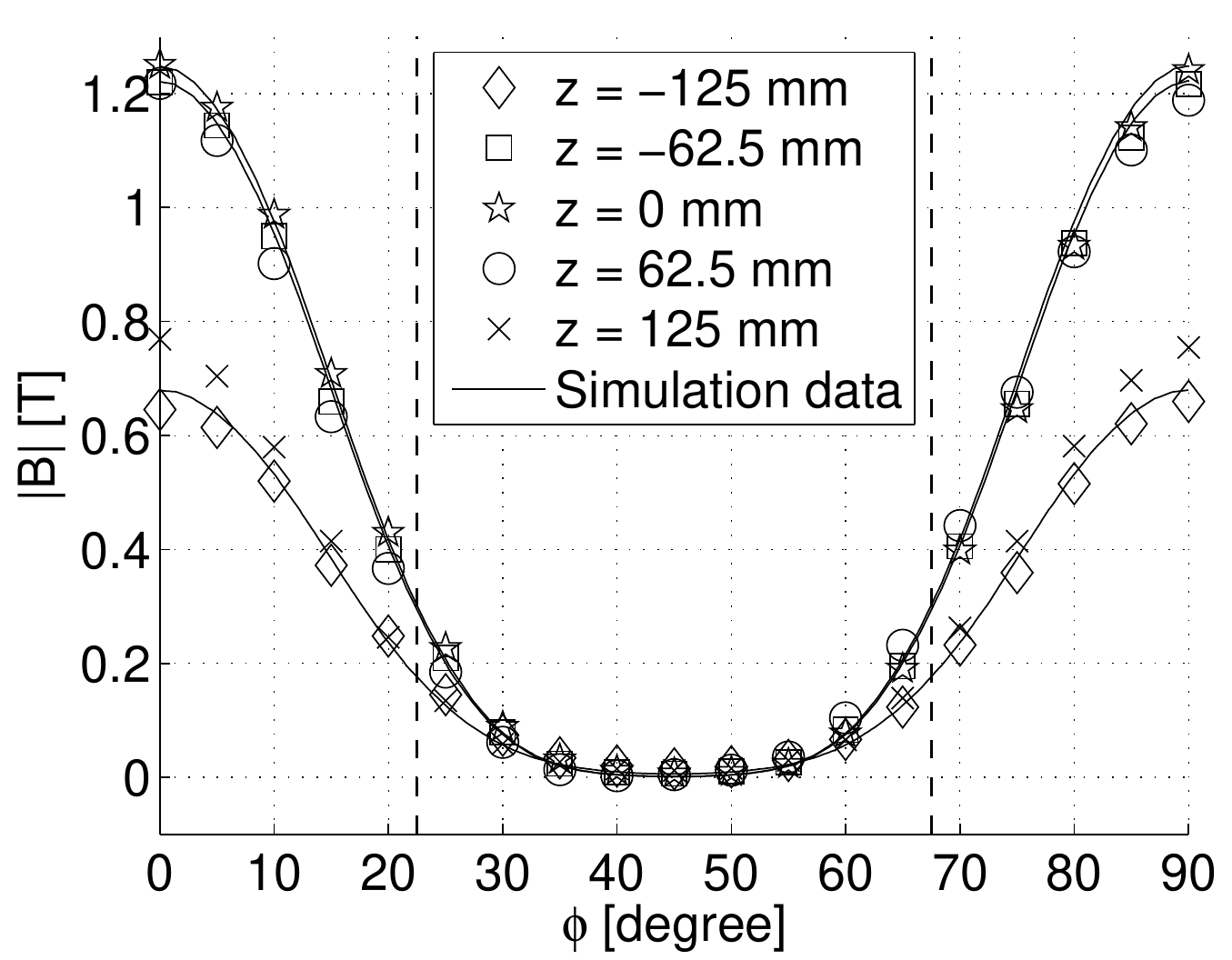}}\hspace{0.2cm}
\subfigure[$|B|$ as function of $z$ in the middle of the air gap, $r=85$ mm.]{\includegraphics[width=.91\columnwidth]{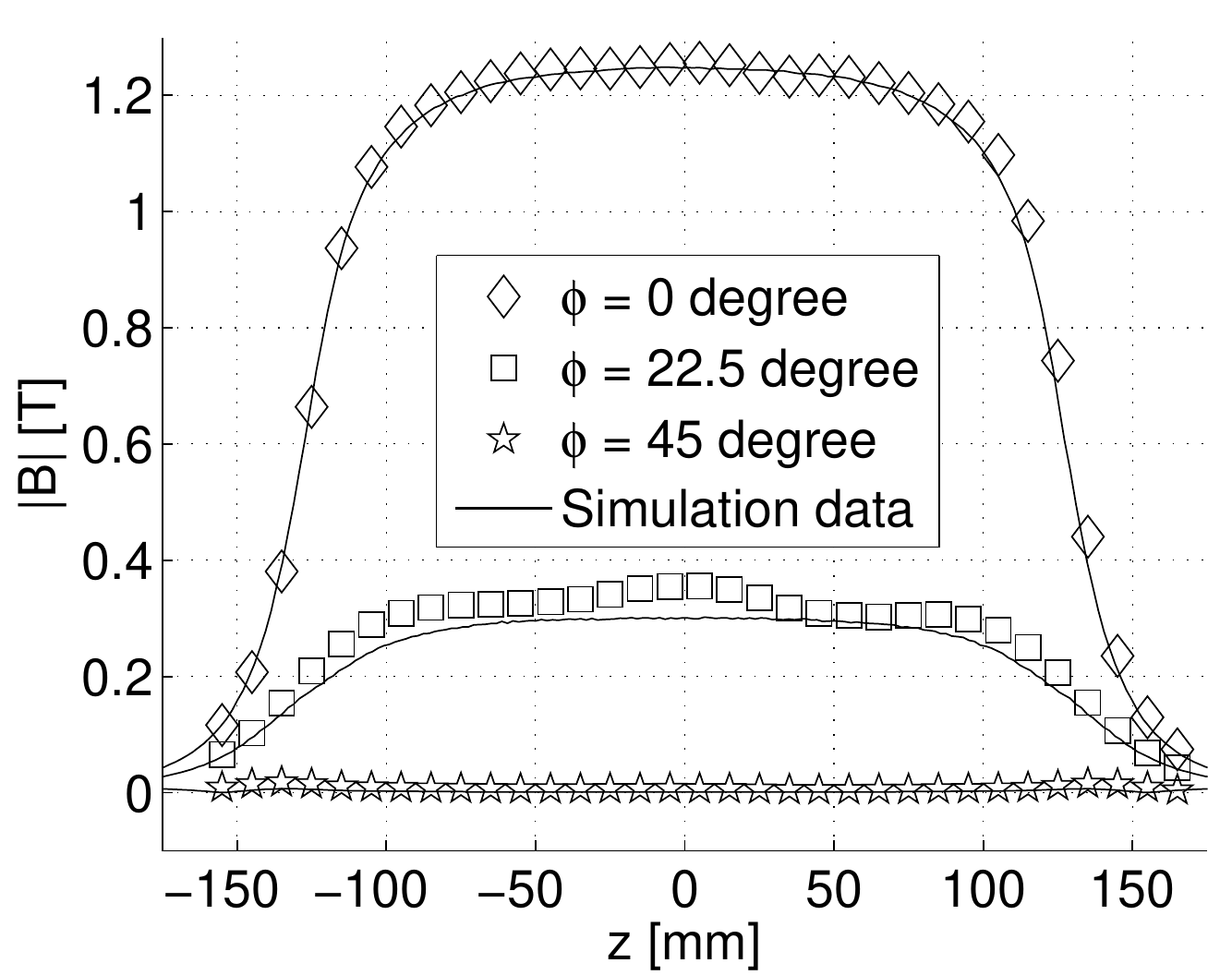}}\hspace{0.2cm}
\subfigure[$|B|$ as function of $r$]{\includegraphics[width=.91\columnwidth]{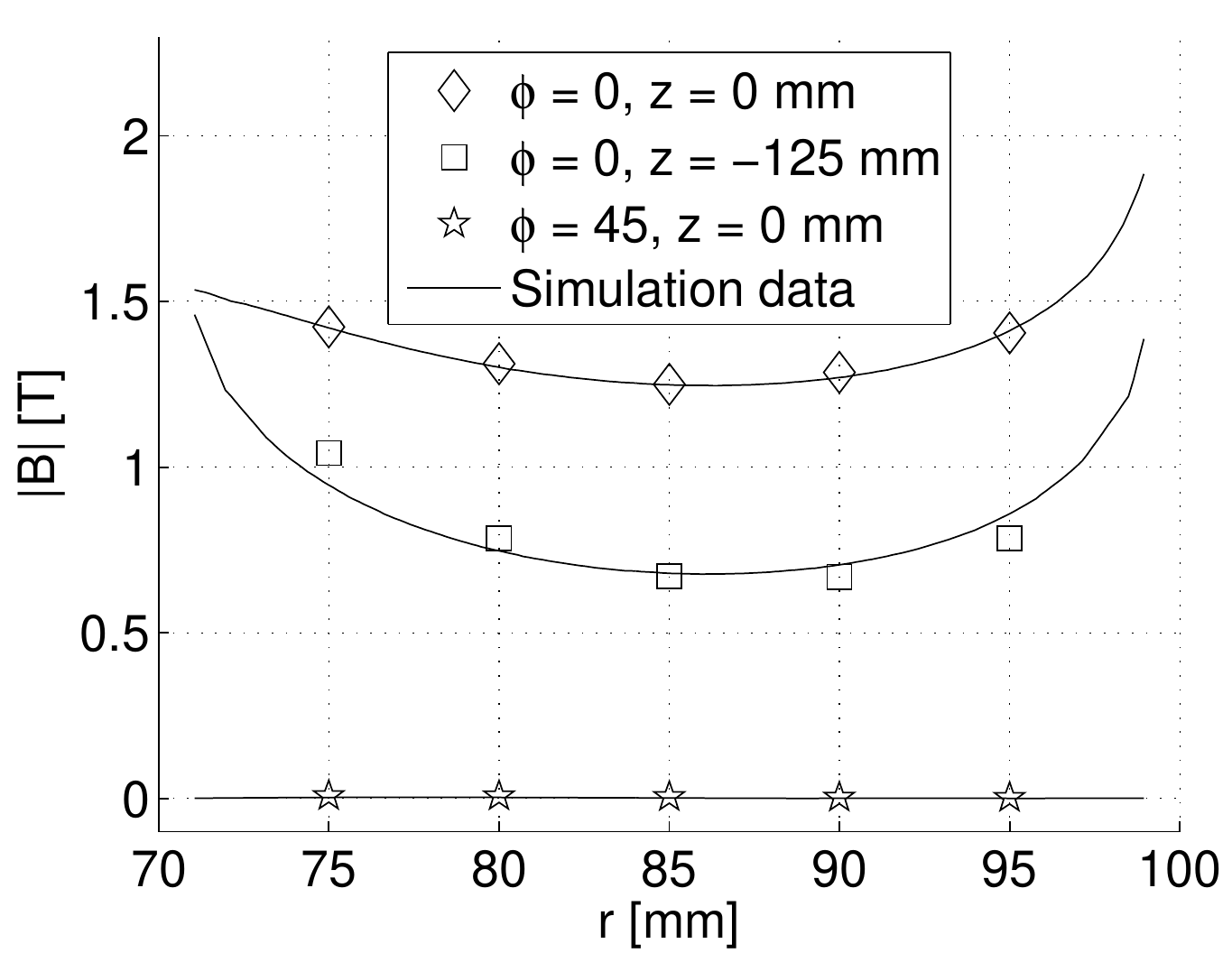}}\\
\caption{Measurements of the flux density as a function of angle $\phi{}$, length $z$, and radius $r$, in the middle of the air gap of the magnet compared with numerical simulations. The dashed vertical lines on Fig. (a) separate the high and low flux density regions.}\label{Fig.Measurements}
\end{figure}

All spatial components of the flux density in the air gap have been measured using a Hall probe (AlphaLab Inc, Model: DCM) as a function of angle, radius and length of the device. A three dimensional simulation of the design has also been performed. The measured flux density was found to be periodic with a period of 90$^\circ$, as expected. The measured flux density for the first 90$^\circ$ and the results of the simulation are shown in Fig. \ref{Fig.Measurements}. An excellent agreement between the simulated and measured flux density is seen.

In the four high field regions the peak flux density is around 1.24 T while it is very close to 0 T in the four low field regions. The gradient between the high and low field regions is quite sharp, but it is clear that the field is not homogeneous in the high field region. However, as the magnetocaloric effect scales with the magnetic field to the power of 0.7 it is preferable to have a zero flux density in the low field region rather than to have part of the flux density gradient in the low field region. Therefore the gradient is concentrated in the high field region. The field is also seen to drop off only at the very ends of the device, i.e. $|z| > 100$ mm. Finally the field is seen to be slightly larger radially near the inner and outer magnet compared to the center of the air gap, but the difference is small and is not expected to have an impact on the performance of the AMR.

\subsection{Performance of the magnet}
The performance of the magnet with regards to magnetic refrigeration can be evaluated using the $\Lambda_\mathrm{cool}$ parameter \cite{Bjoerk_2008}, which is defined as
\begin{eqnarray}
\Lambda_\mathrm{cool} \equiv \left(\langle B_\n{high}^{0.7}\rangle - \langle B_\n{low}^{0.7}\rangle \right)\frac{V_{\mathrm{field}}}{V_{\mathrm{mag}}}P_{\mathrm{field}}~,
\end{eqnarray}
where $V_{\mathrm{mag}}$ is the volume of the magnet(s), $V_{\mathrm{field}}$ is the volume where a high flux density is generated and $P_{\mathrm{field}}$ is the fraction of an AMR cycle that magnetocaloric material is placed in the high flux density volume. Note that $V_{\mathrm{mag}}$ is the volume of permanent magnet material used, excluding any soft magnetic material as the price of this material is in general significantly lower than permanent magnet material. Also, the magnet design presented above has not been optimized with respect to the total weight of the design. More soft magnetic material than needed is present, as the saturation magnetization of the soft magnetic material is not reached. This was done for ease of construction.

Other published magnet designs for magnetic refrigeration devices have a $\Lambda_\mathrm{cool}$ parameter between 0.03 to 0.21 \cite{Bjoerk_2010b}. The magnet designed here has $V_\mathrm{mag} = 7.3$ L, $V_{\mathrm{field}} = 2.0$ L, $\langle B_\n{high}^{0.7}\rangle = 0.91$ T and $\langle B_\n{low}^{0.7}\rangle = 0.15$ T. Assuming $P_{\mathrm{field}}=1$, as is the aim of the device, the design achieves $\Lambda_\mathrm{cool} = 0.21$, thus equaling the best performing magnet published to date. The rotary magnetic refrigeration devices mentioned earlier, Refs. \cite{Tusek_2010}, \cite{Vasile_2006}, \cite{Okamura_2007} and \cite{Zimm_2007} have $\Lambda_\mathrm{cool} = 0.13,\; 0.11,\; 0.21\;\textrm{and}\;0.03$ respectively.

For this particular design the choice of the high and low flux density regions is rather arbitrary and so they could have been chosen to span less than 45 degree. This would lead to a higher value for $\langle B_\n{high}^{0.7}\rangle$ and a lower value of $\langle B_\n{low}^{0.7}\rangle$, but also to a lower value of $V_\mathrm{field}$. It has been verified that $\Lambda_\mathrm{cool}$ attains the highest value for this design when the high and low flux density regions combined span the entire air gap circumference, i.e. as done here.

\section{Conclusion}
The complete process of designing a magnet for use in a magnetic refrigeration device has been described. Two different ways for improving the performance of a magnet design were applied to a concentric Halbach magnet design which was dimensioned and subsequently segmented once the optimal dimensions had been found. The direction of magnetization was also optimized for each of the individual segments. The final design generates a peak value of 1.24 T, an average flux density of 0.9 T in a volume of 2 L using 7.3 L of magnet, and has an average low flux density of 0.08 T. The difference in flux to the power of 0.7 is 0.76 T$^{0.7}$. The working point of the magnets is close the maximum energy density possible. Finally the flux density of the design has been measured and compared with a three dimensional numerical simulation of the design, and an excellent agreement was seen. A magnetic refrigeration device utilizing the magnet is under construction at Ris\o{} DTU.

\section*{Acknowledgements}
The authors would like to acknowledge the support of the Programme Commission on Energy and Environment (EnMi) (Contract No. 2104-06-0032) which is part of the Danish Council for Strategic Research. The authors also wish to thank F. B. Bendixen and P. Kjeldsteen for useful discussions.


\begin{thebibliography}{19}
\expandafter\ifx\csname natexlab\endcsname\relax\def\natexlab#1{#1}\fi
\expandafter\ifx\csname url\endcsname\relax
  \def\url#1{\texttt{#1}}\fi
\expandafter\ifx\csname urlprefix\endcsname\relax\def\urlprefix{URL }\fi

\bibitem{Barclay_1988}
J. A. Barclay, Adv. Cryog. Eng. 33 (1988), 719.

\bibitem{Yu_2003}
B. Yu, Q. Gao, B. Zhang, X. Meng and Z. Chen, Int. J. Refrig. 26 (6) (2003), 622.

\bibitem{Gschneidner_2008}
K. A Gschneidner Jr and V. K. Pecharsky,  Int. J. Refrig. 31 (6), (2008), 945.

\bibitem{Coelho_2009}
A. Coelho, S. Gama, A. Magnus and G. Carvalho, Proc. 3th Int. Conf. on Magn. Refrig. at Room Temp., Des Moines, Iowa, USA (2009), 381.

\bibitem{Dupuis_2009}
C, Dupuis, A. J. Vialle, U. Legait, A. Kedous-Lebouc and D. Ronchetto, Proc. 3th Int. Conf. on Magn. Refrig. at Room Temp., Des Moines, Iowa, USA (2009), 437.

\bibitem{Engelbrecht_2009}
K. Engelbrecht, J. B. Jensen, C. R. H. Bahl and N. Pryds, Proc. 3th Int. Conf. on Magn. Refrig. at Room Temp., Des Moines, Iowa, USA (2009), 431.

\bibitem{Kim_2009}
Y. Kim and S. Jeong, Proc. 3th Int. Conf. on Magn. Refrig. at Room Temp., Des Moines, Iowa, USA (2009), 393.

\bibitem{Sari_2009}
O. Sari, M. Balli, G. Trottet, P. Bonhote, P. Egolf, C. Muller, J. Heitzler and S. Bour, Proc. 3th Int. Conf. on Magn. Refrig. at Room Temp., Des Moines, Iowa, USA (2009), 371.

\bibitem{Tagliafico_2009}
L. Tagliafico, F. Scarpa, G. Tagliafico, F. Valsuani, F. Canepa, S. Cirafici, M. Napoletano and C. Belfortini, Proc. 3th Int. Conf. on Magn. Refrig. at Room Temp., Des Moines, Iowa, USA (2009), 425.

\bibitem{Zheng_2009}
Z. Zheng, H. Yu, X. Zhong, D. Zeng and Z. Liu, Int. J. Refrig. 32 (2009), 78.

\bibitem{Tusek_2010}
J. Tu\v{s}ek, S. Zupan, A. Sarlah, I. Prebil and A. Poredos, Int. J. Refrig. 33 (2) (2010), 294

\bibitem{Bjoerk_2010b}
R. Bj\o{}rk, C. R. H. Bahl, A. Smith and N. Pryds, Int. J. Refrig. 33 (2010{\natexlab{a}}), 437

\bibitem{Vasile_2006}
C. Vasile and C. Muller, Int. J. Refrig. 29 (8) (2006), 1318

\bibitem{Okamura_2007}
T. Okamura, R. Rachi, N. Hirano and S. Nagaya, Proc. 2nd Int. Conf. on Magn. Refrig. at Room Temp., Portoroz, Slovenia (2007), 377.

\bibitem{Zimm_2007}
C. Zimm, J. Auringer, A. Boeder, J Chell, S. Russek and A. Sternberg, Proc. 2nd Int. Conf. on Magn. Refrig. at Room Temp., Portoroz, Slovenia (2007), 341.

\bibitem{Nielsen_2009a}
K. K. Nielsen, C. R. H. Bahl, A. Smith, R. Bj\o{}rk, N. Pryds and J. Hattel, Int. J. Refrig. 32 (6), (2009{\natexlab{a}}), 1478.

\bibitem{Nielsen_2010_priv}
K. K. Nielsen, K. Engelbrecht, C. R. H. Bahl, A. Smith and J. Geyti. Unpublished (2009{\natexlab{b}}).

\bibitem{Bjoerk_2010a}
R. Bj\o{}rk, C. R. H. Bahl and A. Smith, J. Mag. Mag. Mater. 322 (2010{\natexlab{b}}), 133.

\bibitem{Mallinson_1973}
J. C. Mallinson, IEEE Trans. Magn. 9 (4) (1973), 678.

\bibitem{Halbach_1980}
K. Halbach, Nucl. Instrum. Methods 169 (1980).

\bibitem{Bjoerk_2010c}
R. Bj\o{}rk, C. R. H. Bahl, A. Smith and N. Pryds, IEEE Trans. Magn. 47 (6), 1687 (2011).

\bibitem{Bloch_1998}
F. Bloch, O. Cugat, G. Meunier and J. C. Toussaint, IEEE Trans. Magn. 34 (5), 2465 (1998).

\bibitem{Coey_2003}
J. M. D. Coey and T. R. Ni Mhiochain, High Magnetic Fields (Permanent magnets), Edt: F. Herlach and N. Miura, World Scientific, 25 (2003).

\bibitem{Comsol}
COMSOL AB, Tegnérgatan 23, SE-111 40 Stockholm, Sweden.

\bibitem{Standard}
Standard Specifications for Permanent Magnet Materials, Magn. Mater. Prod. Assoc., Chicago, USA. www.intl-magnetics.org. (2000)

\bibitem{Pecharsky_2006}
V. K. Pecharsky and K. A. Gschneidner Jr, Int. J. Refrig. 29 (8) (2006) 1239.

\bibitem{Flemming_2009}
F. B. Bendixen, 2009. Private communication.

\bibitem{Matlab}
Matlab, version 7.7.0.471 (R2008b) (2008).

\bibitem{fminsearchbnd}
J. D'Errico. http://www.mathworks.com/matlabcentral/\\fileexchange/8277, Release: 4 (7/23/06) (2006).

\bibitem{VAC_specifications_2007}
Vacuumschmelze GMBH \&~Co, KG. Pd 002 - Vacodym/Vacomax (2007).

\bibitem{Bjoerk_2008}
R. Bj\o{}rk, C. R. H. Bahl, A. Smith and N. Pryds, J. Appl. Phys. 104 (1) (2008), 13910.
\end{thebibliography}
\end{document}